\documentclass[]{aa}

\usepackage{natbib}
\usepackage{amsmath}
\usepackage{amstext}
\usepackage{graphicx}
\usepackage{txfonts}

\usepackage{verbatim}

\newcommand{\pder}[2]{\frac{\partial#1}{\partial#2}}

\newcommand{\ppder}[2]{\frac{\partial^2#1}{{\partial#2}^2}}

\usepackage{color}

\begin{document} 

\title{Solar energetic particle access to distant longitudes through
  turbulent field-line meandering} 

\author{T. Laitinen
  \inst{1}
  \and
  A. Kopp
  \inst{2}\thanks{previously at the Institut f\"ur Experimentelle und Angewandte Physik,Christian-Albrecht-Universit\"at zu Kiel, Germany}
  \and
  F. Effenberger
  \inst{3,4}
  \and
  S. Dalla
  \inst{1}
  \and
  M.S. Marsh
  \inst{1}\thanks{Now at Met Office, Exeter, UK}
}

\institute{Jeremiah Horrocks Institute, University of Central Lancashire,
  Preston, UK\\
  \email{tlmlaitinen@uclan.ac.uk}
  \and
  Universit\'e Libre de Bruxelles, Service de Physique Statistique et des Plasmas,
  CP 231, 1050 Brussels, Belgium
  \and
  Department of Mathematics, University of Waikato, Hamilton, New
  Zealand
\and
Department of Physics and KIPAC, Stanford University, Stanford, CA, USA
}

\abstract
{ Current solar energetic particle (SEP) propagation models describe
  the effects of interplanetary plasma turbulence on SEPs as
  diffusion, using a Fokker-Planck (FP) equation. However, FP models
  cannot explain the observed fast access of SEPs across the
    average magnetic field to regions that are widely separated in
  longitude within the heliosphere without using unrealistically
  strong cross-field diffusion.}
{ We study whether the recently suggested early non-diffusive phase of
  SEP propagation can explain the wide SEP events with realistic
  particle transport parameters.}
{We used a novel model that accounts for the SEP propagation along
  field lines that meander as a result of plasma turbulence.  Such
a  non-diffusive propagation mode has been shown to dominate the SEP
  cross-field propagation early in the SEP event history. We compare
  the new model to the traditional approach, and to SEP observations.}
{Using the new model, we reproduce the observed longitudinal extent of
  SEP peak fluxes that are characterised by a Gaussian profile with
  $\sigma=30-50^\circ$, while current diffusion theory can only
  explain extents of 11$^\circ$ with realistic diffusion
  coefficients. Our model also reproduces the timing of SEP arrival at
  distant longitudes, which cannot be explained using the diffusion
  model.}
{The early onset of SEPs over a wide range of longitudes can be
  understood as a result of the effects of magnetic field-line random walk in the
  interplanetary medium and requires an SEP transport model that
  properly describes the non-diffusive early phase of SEP cross-field
  propagation.}

{}{}{}


\maketitle

\section{Introduction}\label{sec:introduction}

Solar energetic particles (SEPs) are accelerated up to relativistic
energies during solar eruptions. Their propagation in interplanetary
space is controlled by the large-scale Parker spiral magnetic
field. SEPs can also propagate across the mean field as a result
of large-scale drifts \citep{Marsh2013} and the turbulent magnetic field
fluctuations superposed on the mean field.

SEP propagation in interplanetary space is typically modelled by
solving a Fokker-Planck equation that describes the effect of the
plasma turbulence on the particles as diffusion
\citep{Parker1965,Jokipii1966}. The field-aligned propagation is
affected by small-scale inhomogeneities \citep{Jokipii1966}, whereas
the cross-field propagation is described as diffusion caused by
random-walking field lines
\citep{Jokipii1966,Matthaeus2003,Shalchi2010a}. These theories give
the ratio of the cross-field and field-aligned diffusion coefficients
$\kappa_\perp/\kappa_\parallel\sim 0.01$ near Earth, and these values
are supported by cosmic ray observations
\citep{Burger2000,Potgieter2014} and particle simulations
\citep{GiaJok1999}. Recently, several works have studied the effect of
cross-field diffusion on SEP evolution \citep{Zhang2009,Droge2010,
  He2011, Giajok2012, Qin2013} and showed up to 360$^\circ$
heliolongitudinal extents for wide source regions with onsets within
days of the SEP injection.

Recent multi-spacecraft SEP observations, made with the near-Earth
spacecraft SOHO and ACE and the two STEREO spacecraft, show that SEPs
from a single solar eruption have fast access to a very wide range of
longitudes, even 180$^\circ$ from the eruption location
\citep[e.g.][]{Dresing2012,Richardson2014}. Similar event extents as
measured by longitudinal peak intensity distribution width have been
observed both in gradual and impulsive events
\citep{Wiedenbeck2013,Cohen2014,Richardson2014}, which in traditional
view have very different acceleration origins
\citep[e.g.][]{Reames1999}. Detailed analysis of two SEP
  multi-spacecraft-observed events, with SEP transport fitted with a
  3D FP transport model, favour a small source
region of about~$20^\circ$
   for the SEPs and a high diffusion coefficient ratio
  $\kappa_\perp/\kappa_\parallel\sim 0.1$
  \citep{Dresing2012,Droge2014}. High ratios in the range of $0.1-1$
  have also been inferred in other studies of heliospheric particles
  \citep[e.g.][]{Zhang2003,DwyerEa1997}.

High values of $\kappa_\perp/\kappa_\parallel$ are
  not supported by our current theoretical understanding of SEP
  transport in plasma turbulence, however. At 1~AU, the amplitude of transverse
  magnetic fluctuations, normalised with the background magnetic
  field, $\delta B_\perp/B,$ is observed to be between 0.1 and 0.5
  \citep[e.g.][]{Burlaga1976}, for which full-orbit particle
  simulations suggest $\kappa_\perp/\kappa_\parallel < 0.01$
  \cite{GiaJok1999}. Current theoretical understanding
  also supports much lower values for the ratio \citep[e.g.][]{Pei2010}.

In this paper, we introduce a new model for heliospheric SEP
propagation that is capable of reproducing the fast transport in
longitude that is observed at the beginning of SEP events. Our work is based
on the notion that SEPs spread initially much faster than allowed by a
diffusion description, and that this is a result of field-line meandering
\citep{LaEa2013b}. The cross-field propagation of SEPs along
meandering fields has been analysed previously to study SEP intensity
dropouts \citep{GiaJokMaz2000} and to determine diffusion coefficients
\citep{Kelly2012ApJ}. \citet{LaEa2013b} showed for a uniform magnetic
field that particle propagation across the mean field is initially a
fast non-diffusive process along the meandering field-lines that only
later relaxes to diffusion. Our new approach combines field-line
meandering and time-asymptotic diffusion in a new description of
early-time SEP propagation in the Parker field, the FP+FLRW
  model. We compare the new model to the traditional diffusion
approach (the FP model).

\section{Models}\label{sec:models}

\subsection{FP model}

In the traditional approach, SEP propagation is solved using the
Fokker-Planck equation for the distribution function $f$
  \citep{Roelof1969,Skilling1971,Isenberg1997,Zhang2009}
\begin{multline}
    \frac{\partial f}{\partial t}
+ \left(\mu v\mathbf{b}+\mathbf{V}_{sw}\right)\cdot\nabla f
+ \frac{v}{2L}(1-\mu^2)\frac{\partial f}{\partial \mu} \\
+\left[\frac{\mu(1-\mu^2)}{2}\left(\nabla\cdot\mathbf{V}_{sw}-3\mathbf{b}\mathbf{b}:\nabla\mathbf{V}_{sw}\right)\right] \frac{\partial f}{\partial \mu}
 \\ 
+\left[\frac{1-3\mu^2}{2}
\mathbf{b}\mathbf{b}:\nabla\mathbf{V}_{sw}
-\frac{1-\mu^2}{2}\nabla\cdot\mathbf{V}_{sw}\right]p\frac{\partial f}{\partial
p}\\
= \frac{\partial}{\partial \mu} \left( D_{\mu\mu}\frac{\partial
    f}{\partial \mu}\right)+\nabla\cdot\hat\kappa\nabla f  +Q(\mathbf{r},\mathbf{v},t),\label{eq:FPE}
\end{multline}
where $v$, $p$ and $\mu$ are the particle speed, momentum, and
pitch-angle cosine, $\mathbf{V}_{sw}$ and $\mathbf{b}$ the solar wind
velocity and a unit vector along the local mean magnetic field,
respectively, and $L=-B/ ( \partial B/\partial s)$ the focusing length
of the particles, with $s$ the arc length along the field line, and $Q$
is the injection function. The terms of the order
  $(dV_{sw}/dt)/v$ \citep{Isenberg1997} are typically neglected in SEP transport studies as small compared to the other terms \citep[e.g.][]{Zhang2009,Droge2010}, and they vanish for the constant radial solar wind considered in our study (see Sect.~\ref{sec:turbulence-model}).

The pitch-angle diffusion coefficient $D_{\mu\mu}$ and the cross-field
diffusion coefficient $\kappa_\perp$ are given as
\citep{Jokipii1966,Matthaeus2003}
\begin{eqnarray}
    D_{\mu\mu}&=&\frac{\pi\Omega^2}{v|\mu| B^2}(1-\mu^2)
  S_\parallel\left(-(r_L\mu)^{-1}\right) \label{eq:dmumu}\\
  \kappa_\perp&=&\frac{ a_{NLGC}^2 v^2}{3B^2}\int d^3 k
  \frac{S(\mathbf{k})}{v/\lambda_\parallel+k_\perp^2\kappa_\perp+k_\parallel^2\kappa_\parallel}\label{eq:nlgc}
,\end{eqnarray}
where $S(\mathbf{k})$ is the turbulence power spectrum, $S_\parallel$
the spectrum of slab turbulence (see Sect.~\ref{sec:turbulence-model}), $\Omega$ and $r_L$ are the
particle Larmor frequency and radius, respectively, the free
parameter $a_{NLGC}=1/\sqrt{3}$ \citep{Matthaeus2003}, and the
parallel mean free path
\begin{equation}
  \lambda_\parallel=\frac{3v}{4}\int_0^1 d\mu\frac{(1-\mu^2)^2}{D_{\mu\mu}}.\label{eq:lambdapar}
\end{equation}
The field-aligned spatial diffusion tensor is given as
\begin{equation}
  \hat\kappa=\left(
  \begin{array}{ccc}
    \kappa_\perp & 0 & 0\\
    0 & \kappa_\perp & 0\\
    0 & 0 & 0
  \end{array}\right),
\end{equation}
that is, without parallel diffusion, because this is accounted for with
$D_{\mu\mu}$.

The changes in SEP energy during propagation in the
  interplanetary space are given by the terms in the third line of
  Eq.~(\ref{eq:FPE}). They include momentum changes that are
due to the diverging
  solar wind, and in an approximate way, the effect of drifts
  on the particle momentum \citep{Dalla2015}. However, the equation
  does not contain the drifts in its convective term (second term in
  the first line), which means that the transport picture presented by
  Eq.~(\ref{eq:FPE}) is incomplete. The effect of latitudinal drifts
  on the particle energy was recently discussed by \citet{Dalla2015},
  who found that in 100~hours a 10~MeV proton can lose almost half of
  its energy through drifting along the
  $\mathbf{V}_{sw}\times\mathbf{B}$ electric field and adiabatic
  deceleration. An improved approach, the
  drift kinetic theory, introduces drifts within the convective term
  \citep[e.g.][]{leRouxWebb2009,WebbEa2009}.

The energy changes that are due to drifts and the diverging solar wind
may be significant when performing a detailed fitting of the SEP event
onset and the decay phase \citep[e.g.][]{Ruffolo1995,
  Dalla2015}. However, we here consider neither, but only
the first ten hours of the event. For simplicity, the
energy change term is therefore neglected in this study. Because the spatial
  transport associated with drift is small for a 10 MeV proton within
  the early phase of an SEP event \citep{Dalla2015}, drifts are
  neglected within the convective term.

Particle diffusion parallel and across the mean field direction, as
understood by current theories, are not independent of each other. The
cross-field diffusion coefficient in Eq.~(\ref{eq:nlgc}) depends on
the parallel diffusion coefficient, resulting in compound diffusion
\citep[e.g.][]{Kota2000,Qin2002apjl,Matthaeus2003}. The pitch angle dependence of the cross-field
diffusion coefficient has also been discussed recently, with suggestions for using $\kappa_\perp\propto
|\mu|$ \citep[e.g.][]{QinShalchi2014} or $\kappa_\perp\propto r_L$,
the particle Larmor radius \citep{Droge2010}. While the suggested
different forms of pitch angle dependence of $\kappa_\perp$ influence
particle cross-field propagation \citep[e.g.][]{Strauss2015}, it has
not been clearly established which type of dependence best represents
particle behaviour in turbulent fields. We therefore use
the conventional pitch-angle independent form of $\kappa_\perp$
here.

We solved Eq.~(\ref{eq:FPE}) using stochastic differential equations
\citep[SDEs, e.g.][]{Gardiner2009}, which solve the equations using
  pseudo-particles. The method is briefly
  described in Appendix~\ref{sec:stoch-diff-equat}, further details
  are given in \citet{Kopp2012}. We used enhanced pitch angle
  scattering across $\mu=0$ by using a pitch angle diffusion
  coefficient of the form
\begin{equation}\label{eq:resgap}
  D_{\mu\mu}=\kappa_0\left(\left|\mu\right|^{q-1}+H\right)\left(1-\mu^2\right),
\end{equation}
suggested by \citet{BeeckWibberenz1986}, where $\kappa_0$ contains the
$\mu$-independent terms in Eq.~(\ref{eq:dmumu}), $q$ is the spectral
index of the slab turbulence, and $H=0.1$ is a parameter that enhances
scattering between pitch angle hemispheres.  The FP equation is
solved in a Parker spiral magnetic field with the magnitude $B$ given as
\begin{equation}
  B(r)=B_0\left(\frac{r_0}{r}\right)^2 \sqrt{\frac{r^2+a^2}{r_0^2+a^2}},
\end{equation}
where $B_0=5$~nT is the magnetic field at heliocentric distance
$r_0=1$~AU, and $a=V_{sw}/(\Omega_\odot\sin\theta)$, where
$V_{sw}=400$~km/s, $\Omega_\odot=2.8631\cdot10^{-6}$~rad/s is the
solar rotation rate and $\theta$ the colatitude. 

\subsection{FP+FLRW model}\label{sec:fp+flrw-model}

\citet{LaEa2013b} demonstrated that the early-time particle transport
across the mean magnetic field is dominated by the particles
propagating along meandering field lines. With respect
  to the mean magnetic field direction, the early-time cross-field
  propagation of the particles is therefore only weakly stochastic. At later times,
\citet{LaEa2013b} showed that the cross-field extent of the particle
population begins to widen, resulting in an asymptotic diffusive
behaviour.

In this study, we used an FP+FLRW model that explicitly
  introduces the random walk of field lines across the Parker spiral
  magnetic field and particle propagation along these random-walking
  field lines. A similar model was successfully used by
  \citet{LaEa2013b} for Cartesian geometry. In addition
  to propagating particles along meandering field lines, the particles
  are diffused across them with the time-asymptotic cross-field
  particle diffusion coefficient $\kappa_\perp$ to facilitate the
  late-time widening of the particle population. \citet{LaEa2013b}
  showed that such a model reproduces both the initial wide
  extent of the SEP event (where the FLRW dominates) and the
  asymptotic diffusive behaviour at later times (where the particle
  diffusion dominates) well and that is significantly better than the FLRW or
  FP on their own.

In our model, the field line random walk is facilitated by the
  turbulent interplanetary magnetic field, which is described by a
  power spectrum. A mix of 2D and slab modes in turbulent magnetic
  field gives rise to diffusive spreading of field lines
  \citep[][]{Matthaeus1995}. This allows us to describe the path of a
  particle following a meandering field line using SDEs, with
\begin{equation} \label{eq:fldiff}
  dr_\perp(r_\parallel)=\sqrt{2 D_{FL}(r_\parallel)dr_\parallel}W_\perp,
\end{equation}
where $dr_\perp$ is the displacement across the Parker field direction
for advance $dr_\parallel$ along the field, $D_{FL}$ is the field-line
diffusion coefficient, and $W_\perp$ a Gaussian random number with
zero mean and unit variance. The random-walking path is
  calculated for each simulated particle before the pseudo-particle
  trajectory is integrated. The pseudo-particles are propagated along
  and diffused from the meandering field lines according to
  Eq.~(\ref{eq:FPE}), with cross-field diffusion given by
  Eq.~(\ref{eq:nlgc}), pitch angle diffusion as given in Eqs.
  (\ref{eq:dmumu}) and~(\ref{eq:resgap}), and focusing, where we used
  the Parker field geometry when calculating the focusing length.

The field-line diffusion coefficient was obtained as presented in
\citet{Matthaeus1995}, for composite slab and 2D turbulence. The
contribution from the slab component is proportional to
$S_\parallel(k_\parallel=0)$, which in our turbulence model (see
below) vanishes. We therefore only considered the 2D contribution to
the field-line diffusion, giving \citep{Matthaeus1995}
\begin{equation}
  D_{FL}=\left\{\frac{\int d^2 \mathbf{k}_\perp S_{xx}(\mathbf{k}_\perp) /
    k_\perp^2}{B^2}\right\}^{1/2}.\label{eq:dfl}
\end{equation}
It should be noted that particle propagation as a beam along
  field lines that random-walk according to the field-line diffusion
  coefficient in Eq.~(\ref{eq:dfl}) causes transport across the mean
  field much faster than diffusion with the coefficient given by
  Eq.~(\ref{eq:nlgc}). The latter aims to describe the cross-field
  spreading of particles at late times, when the pitch-angle
  distribution of the particles is isotropic. The two descriptions
  converge for a particle beam in turbulence with vanishing slab
  component, $a_{\textrm{NLGC}}=1$ and a scatter-free velocity
  correlation $\left<v_z(0)v_z(t')\right>=v_z^2$ in Eq.~(3) of
  \citet{Matthaeus2003}. The two descriptions of particle
  cross-field propagation are therefore consistent with each other in the
  appropriate limit.

Consistent with the SDE approach in the traditional model, the effect
of the cross-field step length is not considered in the particle
propagation time. This affects the accurate timing of particle
propagation \citep[e.g.][]{Strauss2014}.

\begin{figure*}
  \begin{center}
    \includegraphics[width=0.98\columnwidth]{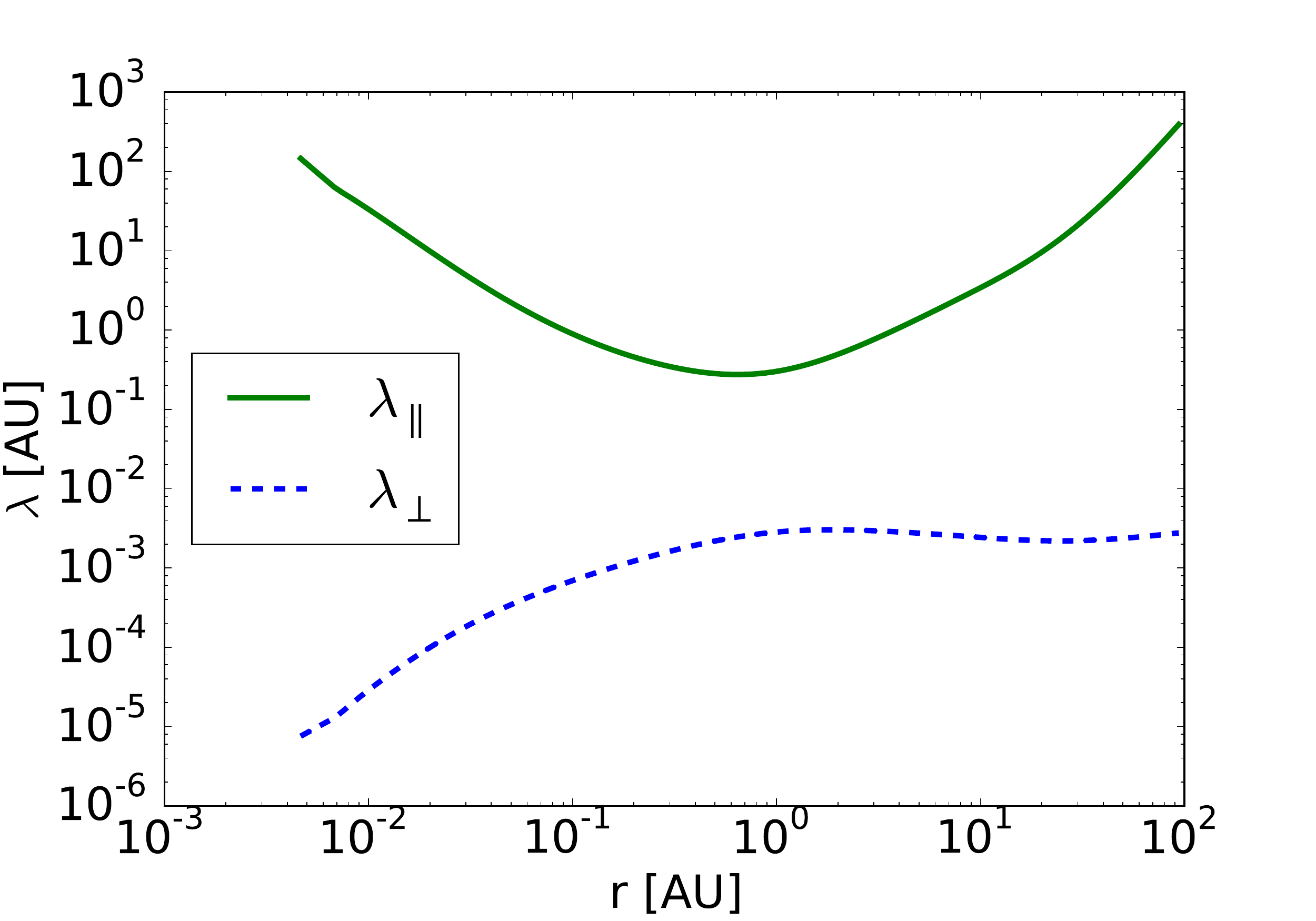}\hfill
    \includegraphics[width=0.98\columnwidth]{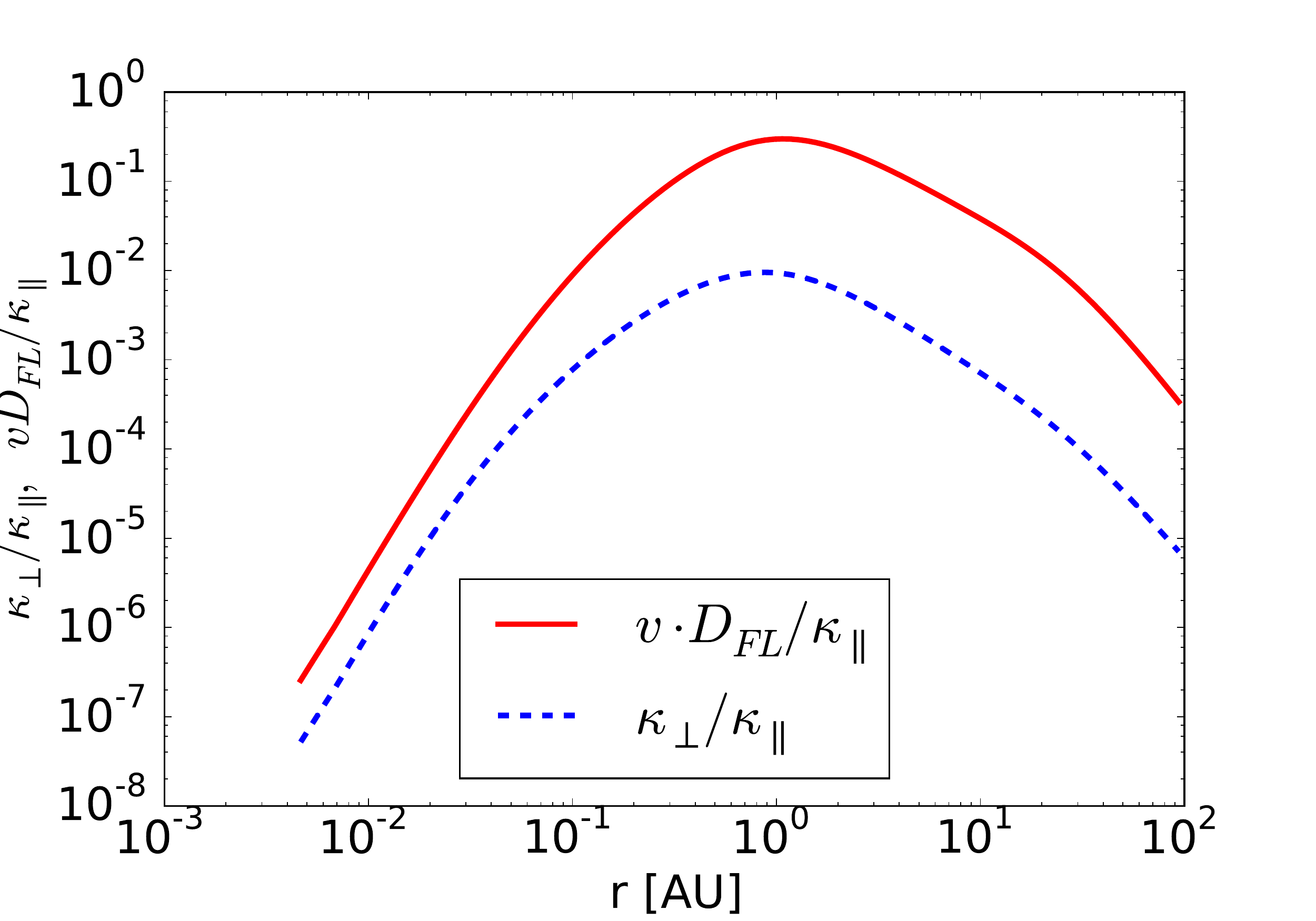}
  \end{center}
  \caption{Left: the parallel and perpendicular scattering mean free
    paths for 10 MeV protons obtained with
    Eqs.~(\ref{eq:dmumu})-(\ref{eq:lambdapar}) and
    $\lambda_\perp=3\kappa_\perp/v$. Right: Diffusion coefficient
    ratios. For the red curve, the cross-field diffusion coefficient
    is obtained as $\kappa_\perp=v D_{FL}$, representing a cross-field
    diffusion coefficient that is due to field-line meandering of unscattered
    particles.}\label{fig:kappas}
\end{figure*}

\subsection{Turbulence model}\label{sec:turbulence-model}

To compare the FP and FP+FLRW models, it is essential that
  $D_{\mu\mu}$, $\kappa_\perp$ and $D_{FL}$, as defined in
  Eqs.~(\ref{eq:dmumu}),~(\ref{eq:nlgc}) and~(\ref{eq:dfl}), are
  consistent with each other. We ensured this by deriving them for a
  simple interplanetary turbulence model. We considered only wave modes
parallel (slab) and perpendicular (2D) to the mean magnetic field
\citep{Gray1996} and defined the power spectrum as
\begin{equation}
  \label{eq:spectrum}
  S(\mathbf{k})\equiv S(\mathbf{k},r_0)=S_\perp(k_\perp)\delta(k_\parallel)+S_\parallel(k_\parallel)\delta(\mathbf{k}_\perp),
\end{equation}
where $k_\parallel$ and $k_\perp=\left|\mathbf{k}_\perp\right|$ are
magnitudes of the wave number. The spectra are
given as broken power laws
\begin{eqnarray}
  \label{eq:parspec}
  S_\parallel(k_\parallel)&=&\begin{cases}
    C_\parallel L_\parallel \delta B_\parallel^2 & L_0^{-1}<k_\parallel<L_\parallel^{-1} \\
\;&\;\\
    \displaystyle{\frac{C_\parallel L_\parallel \delta B_\parallel^2}{\left(L_\parallel k_\parallel\right)^{5/3}}} & k_\parallel\geq L_\parallel^{-1}
  \end{cases}\\
\;&&\;\nonumber\\
\mbox{and} && \nonumber \\
\;&&\;\nonumber\\
  \label{eq:perpspec}
  2\pi k_\perp S_\perp(k_\perp)&=&\begin{cases}
    C_\perp L_\perp \delta B_\perp^2 & L_0^{-1}<k_\perp<L_\perp^{-1} \\
\;&\;\\
    \displaystyle{\frac{C_\perp L_\perp \delta B_\perp^2}{ \left(L_\perp k_\perp\right)^{5/3}}} & k_\perp\geq L_\perp^{-1},
  \end{cases}
\end{eqnarray}
where $L_\parallel$ and $L_\perp$ are the spectral breakpoint scales
with a value of 0.007~AU used for both scales, consistent with
  solar wind observations \citep[e.g.][and references
  therein]{TuMarsch1995}. The largest scale in the turbulence model,
$L_0(r)$, is taken to be $r$, a natural choice for scaling in
spherically expanding solar wind. The turbulence amplitude $\delta
B^2$ is related to the normalisation factors $C_\parallel$ and
$C_\perp$ to give $\delta
B_{\parallel,\perp}^2=2\int
S_{\parallel,\perp}(\mathbf{k})d\mathbf{k}$. The turbulence amplitudes
are normalised to give the parallel scattering mean free path
$\lambda_\parallel=0.3$~AU for a 10~MeV proton at 1~AU heliocentric
distance \citep[e.g.][]{Palmer1982}, and $\delta B^2_\parallel:\delta
B^2_\perp=20\%:80\%$ \citep{Bieber1996}, giving $\delta
  B^2/B^2=0.04$ at 1~AU, in line with interplanetary turbulence
  observations \citep[e.g.][]{Burlaga1976}.

  The plasma turbulence evolves as the solar wind propagates
    from the Sun to the interplanetary medium
    \citep[e.g.][]{TuMarsch1995}. The amplitude of the fluctuations
    changes as a function of radial distance as the properties of the
    background plasma change. The turbulence also evolves
    non-linearly, with energy cascading towards smaller scales. The
    radial and spectral evolution of turbulence in the heliosphere
    have been studied in several works over several decades. Many of
    the works, however, considered isotropic or slab turbulence
    \citep[e.g.][]{TuPuWei1984, Tu1987, Zhou1990swturbtrans, VLF2003}
    instead of the more realistic anisotropic geometry, or the
    composite mode used in this and many other recent SEP transport
    studies. Some formulations exist for spectral transport separately to
    small parallel and perpendicular scales    \citep[e.g.][]{Cranmer2005,Laitinen2005}. These works, however,
    have not been considered with respect to particle transport
    coefficients. The recent detailed heliospheric turbulence
    transport model by \citet{Zank2012} provides the
    evolution of the total turbulence energy and correlation lengths
    instead of the turbulence spectrum, and the application of their
    model to SEP transport modelling is a challenging task that
has so far not
    been attempted.

    Developing an SEP transport model that    considers the spectral evolution of composite turbulence in the
    heliosphere is therefore an undertaking that has not been considered before, and it
    warrants a separate study. In this work, we take a simpler approach
    and only consider the radial evolution of the turbulence. We used
  the WKB description \citep{Richter1974,TuPuWei1984} for the slab
  and 2D components. For simplicity, we did not consider the evolution
  of the spectral shape nor the wave refraction, and we kept the wave
  components in their pure slab and 2D geometries.  Furthermore, we did not
  consider the change in $k$ that is due to the varying wave phase speed because  assuming a constant inertial-frame wave phase speed agrees well
  with observations \citep{VLF2003}. With these simplifications, the
  turbulence evolves as
\begin{eqnarray}
S_{\parallel,\perp}(k_{\parallel,\perp},r)&=&S_{\parallel,\perp}(k_{\parallel,\perp},r_0)\left[\frac{V_{r0}\, v_{Ar}}{V_r\, v_{Ar0}}\right]^2
\left(\frac{n_e}{n_{e0}}\right)^{1/2} \nonumber \\
&\equiv& S_{\parallel,\perp}(k_{\parallel,\perp},r_0)W(r),\label{eq:WKB}
\end{eqnarray}
where $v_{Ar}$ is the radial component of Alfv\'en velocity,
$V_r=V_{sw,r}+v_{Ar}$ and $n_e$ the solar wind electron number
density, with the subscript 0 denoting the values at reference distance
$r_0$.  We used a constant solar wind velocity $V_{sw,r0}$ and electron
density $ n_e(r)=n_{e0}\;r_0^2/r^2$.  Using these, we find
\begin{equation}
  W(r)=\left(\frac{r_0}{r}\right)^3\left(\frac{V_{sw,r0}+v_{A0}}{V_{sw,r0}+\frac{r_0}{r}v_{A0}}\right)^2,
\end{equation}
which is consistent with the observed approximately $\propto r^{-3}$
trend in interplanetary space \citep[e.g.][]{Bavassano1982JGR}. We here used $V_{sw,r0}=400$~km/s and $v_{a,r0}=30$~km/s to
represent the values at $r_0=1$~AU.

It should be noted that our model for the field-line random walk
  differs from that of \citet{Giacalone2001}, where the field-line
  random walk is caused by the motion of the magnetic field footpoints
  at the Sun that are due to supergranulation. The \citet{Giacalone2001} model
  provides an explanation for the source of the field-line random walk at
  the Sun, but it does not allow for turbulence evolution in
  interplanetary space, which further modifies the turbulence as the
  magnetic fields are convected with the solar wind \citep[e.g.][and
  references therein]{Bruno2005}. In our model, the field-line
  meandering is derived from the turbulence spectrum and the corresponding
  field line diffusion coefficient, with no limiting assumptions on
  the interplanetary evolution of the turbulence.

\subsection{Transport parameters}

After specifying a turbulence model, we now calculate the
  particle transport parameters that are fully consistent with the
  properties of the turbulence for both the FP and FP+FLRW models.
  The transport parameters are shown in Fig.~\ref{fig:kappas} for 10
  MeV protons. In the left panel we show the particle transport
  parameters as parallel and cross-field mean free paths, which are
  the same for both the FP and the FP+FLRW model, and the right
  panel shows the ratio of the cross-field and parallel diffusion
  coefficients, $\kappa_\perp/\kappa_\parallel$. The values of the
  particle diffusion coefficients are broadly similar to those
  used in other studies \citep[e.g.][]{Zhang2009,Pei2010}.  

At small heliocentric distances, our model gives high values for
  the parallel mean free path (left panel of Fig.~\ref{fig:kappas}),
  allowing the particles to propagate along the field lines
  essentially without scattering from the Sun to 0.4~AU. A similar
  transport regime is also present in models that use a constant radial
  diffusion coefficient to parametrise the parallel diffusion
  coefficient \citep[e.g.][]{Zhang2009,Droge2010, He2011, Giajok2012,
    Qin2013}.

In addition to
  $\kappa_\perp/\kappa_\parallel$, we show also the ratio for
  particles that propagate as a beam along the meandering field line,
  $v\cdot D_{FL}$, in the right panel of Fig.~\ref{fig:kappas}. As discussed in Sect.~\ref{sec:fp+flrw-model},
  $v\cdot D_{FL}$ is always larger than $\kappa_\perp$. In addition,
  as shown by \citep{Fraschetti2011perptimetheory} and
  \citet{Fraschetti2012perptimesims}, for typical interplanetary
  turbulence parameters the rate of decoupling of particles from their
  field lines is always slower than the spreading of particles across
  the mean field. Thus, at early times the SEP propagation along
  random-walking field lines dominates over both the asymptotic
  perpendicular diffusion and the particle decoupling from the field.
  
\begin{figure*}
  \begin{center}
    \includegraphics[width=0.98\columnwidth]{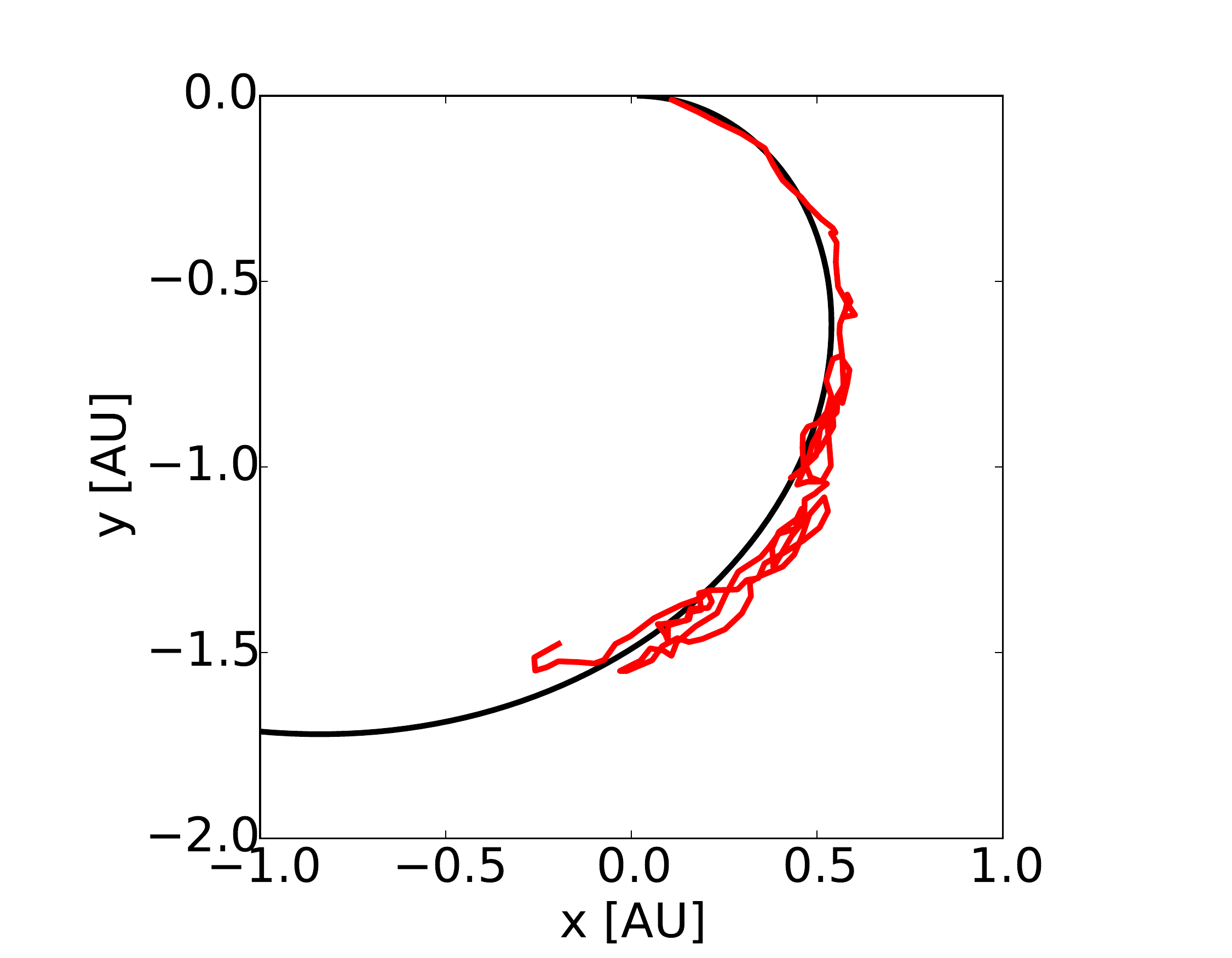}\hfill
    \includegraphics[width=0.98\columnwidth]{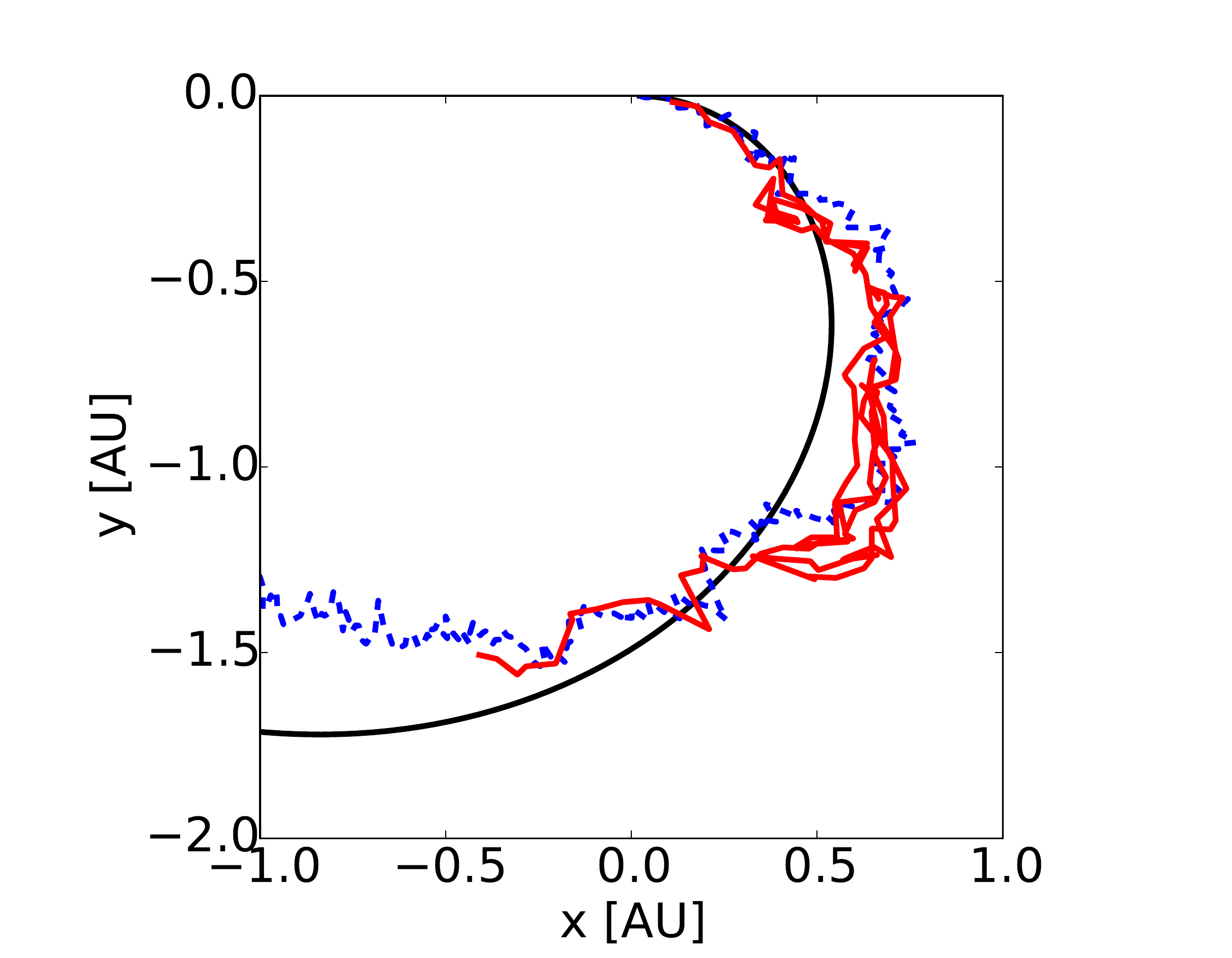}
  \end{center}
  \caption{Sample path (red curve) of a 10 MeV proton injected at
    the Sun (located at origin) with the FP
    (left panel) and FP+FLRW model (right panel), respectively. The
    black curve shows the Parker spiral field, while the dashed blue
    curve shows a meandering field line in the FP+FLRW model (right
    panel).}\label{fig:paths}
\end{figure*}

The cross-field mean free path for both particles and field
  lines is very small close to the Sun, of the order of $10^{-5}$~AU.
  The value is deceptive, however: a low value close to the
  Sun can result in large longitudinal spread. In radial geometry, the
  field-line diffusion coefficient can be written in terms of a
  longitudinal diffusion coefficient, $D_{FL}=\left<\Delta
    x^2\right>/(2r)=r^2\left<\Delta\phi^2\right>/(2r)\equiv r^2
  D_{FL\phi}$. In a radial geometry, a constant longitudinal
  diffusion coefficient is therefore represented by a diffusion coefficient that
  is proportional to $r^2$. For our turbulence model, we have
  $D_{FL\phi}=(40^\circ)^2$/AU close to the Sun, and
  $D_{FL\phi}=(10^\circ)^2$/AU at 1~AU.

It should be noted that a simple parametrisation with
  $\kappa_\perp/\kappa_\parallel$ a constant or $\propto r$ does not
  reproduce the behaviour of the diffusion coefficients obtained with
  a full model of the turbulence, such as the one used in this study
  and in \citet{Pei2010}. The radial evolution of turbulence and its
  effect on the diffusion coefficients will be discussed in more
  detail in a separate study.

\section{Results and discussion}\label{sec:results}

We here study a simple injection profile,
\begin{eqnarray*}
  \label{eq:inj}
  Q(r,\theta,\phi,t)&=&\delta(r-1 \mathrm{r}_\odot) \delta(\theta-\pi/2) \cdot \\
  &&\delta(\mu-1)\delta(E-E_0)\delta(\phi) \delta(t-t_0),
\end{eqnarray*}
for $E_0=10$~MeV protons and trace the propagation of
  pseudo-particles through interplanetary space according to the
model equations. The resulting profile can also be convolved
  with different injection profiles to produce response for time- and
  longitude-extended injections, which will be a subject of a future
  full parameter study.

  The propagation of the particles in FP and FP+FLRW models is
    shown in Fig.~\ref{fig:paths} as a 2D projection of the 3D
    orbits. The red curves depict the path of a simulated pseudo-particle,
    and the black curves show the Parker spiral. In the right panel, the
    blue dashed curve depicts the diffusively meandering field
    line. In both models, the particles scatter along the field line
    and diffuse away from it in the cross-field direction. In the
    FP+FLRW model, however, the field line itself meanders away from
    the Parker Spiral field. As discussed above, the diffusion of the
    field line is faster than the diffusion of a particle across the
    field line, thus the particle orbit tends to follow the meandering
    field line (blue dashed curve in Fig.~\ref{fig:paths}), and as a
    result spreads faster across the mean Parker spiral field than in
    the FP case.

\begin{figure*}
\begin{center}
  \includegraphics[width=.8\textwidth]{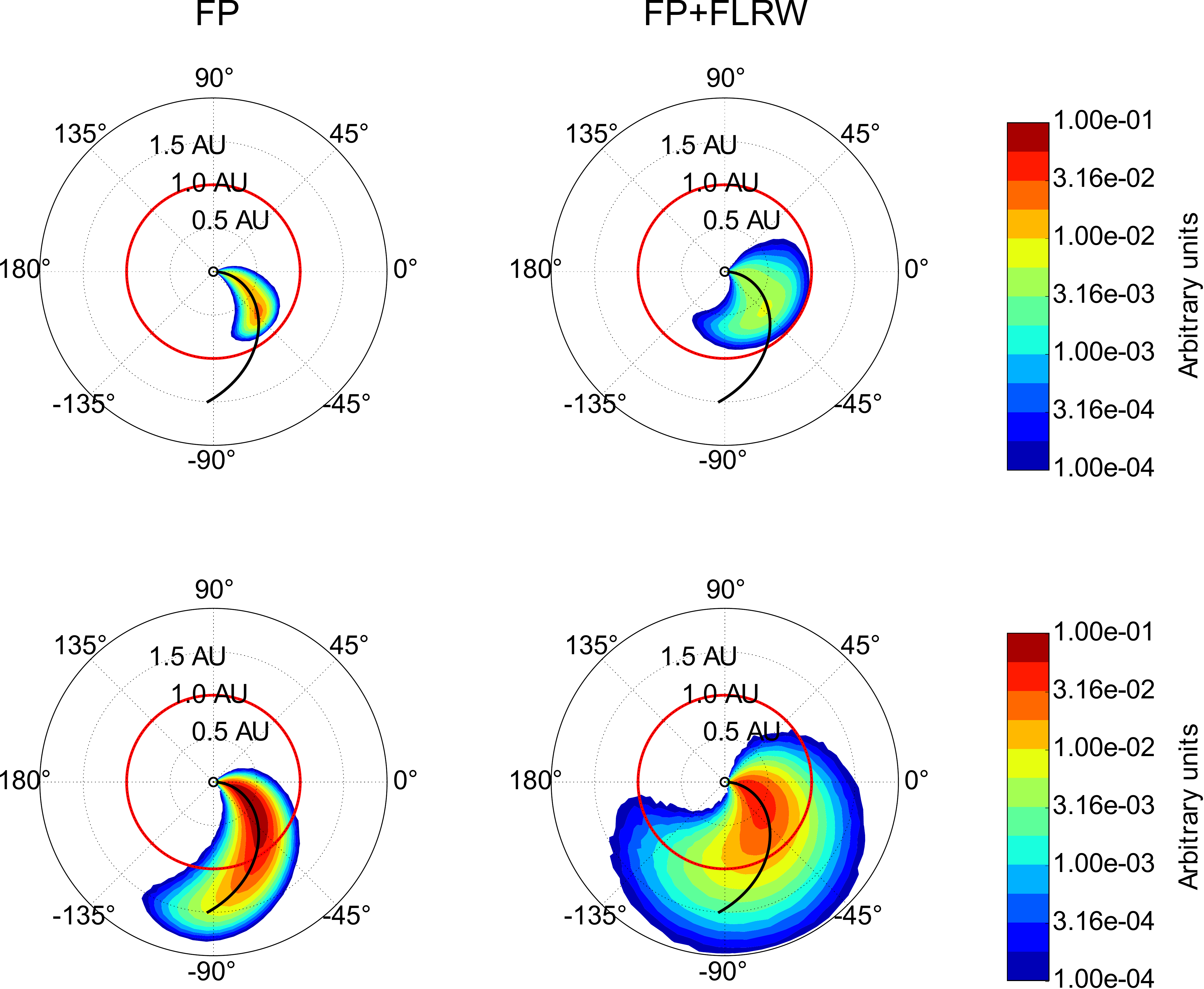}
\end{center}

  \caption{SEP particle density in arbitrary units for particles between
  latitudes $-10^\circ$ and $10^\circ$, one hour (top row) and three hours
  (bottom row) after an impulsive injection at 1 solar radius and
  longitude $\phi=0^\circ$. Left panels show the result of the FP
  model, while the right panels show the FP+FLRW model. The black
  curve shows the Parker field-line starting at $\phi=0^\circ$. The
  field line reaches a radial distance of 1~AU (red circle) at
  $\phi=-62^\circ$.}\label{fig:parker}
\end{figure*}

In Fig.~\ref{fig:parker} we compare the particle distributions
predicted by the diffusion (FP) (left) and our new FP+FLRW (right)
model, one (top row) and three (bottom row) hours after injection. The
distribution contours are calculated for particles near the ecliptic
plane, between latitudes $\pm 10^\circ$.  In the FP model, particles
diffusively spread in longitude as a function of time, and as they are
focused rapidly away from the Sun in the diverging magnetic field, the
most significant spreading of the particles takes place at distances
$>0.5$~AU. In the FP+FLRW model, however, the particles have a much
wider distribution in longitude and show significant spreading close
to the Sun.

\begin{figure*}
\begin{center}
\includegraphics[width=.8\textwidth]{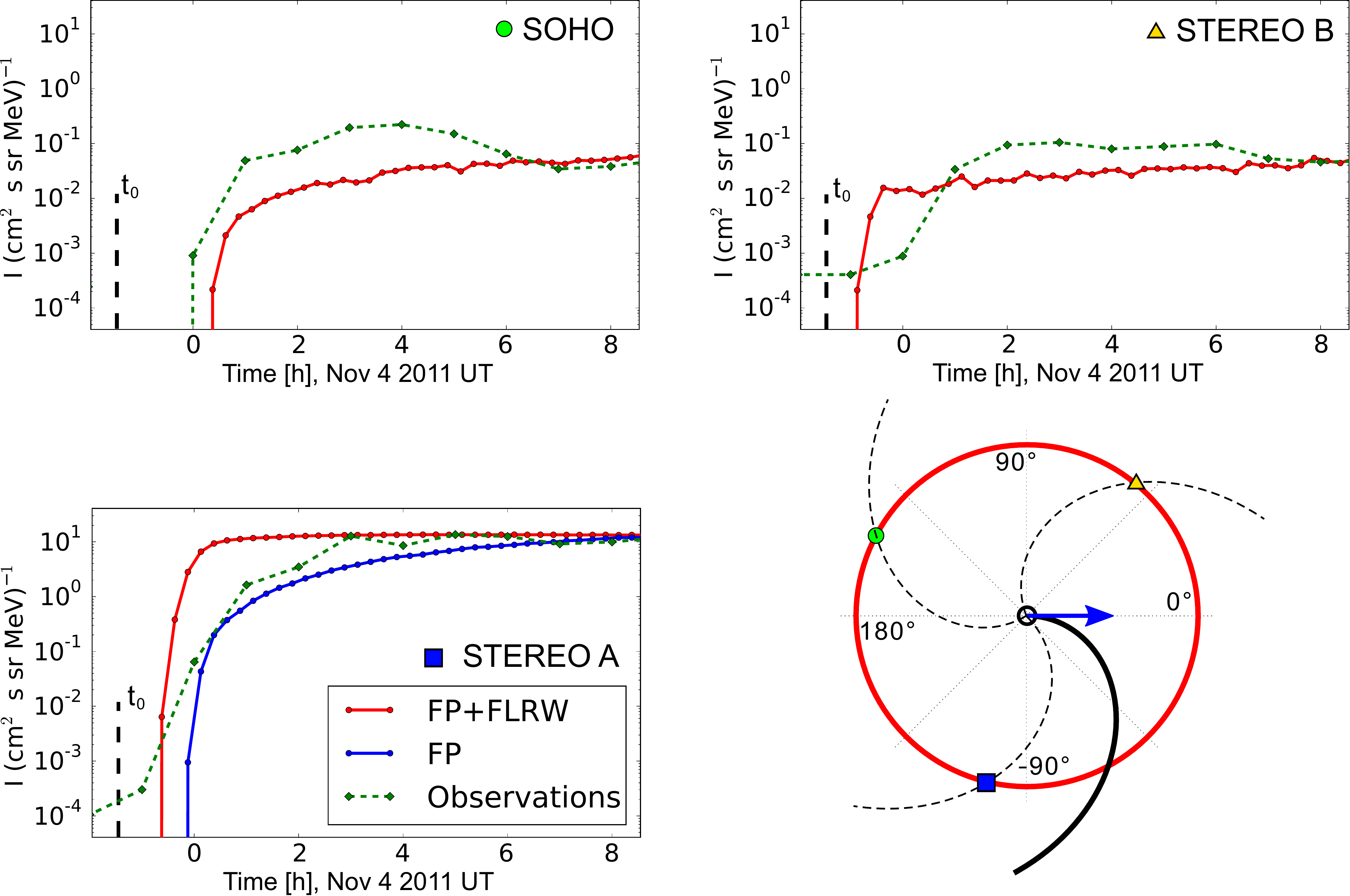}
\end{center}
\caption{SEP proton intensities as a function of time at
    1~AU. The green dashed curve shows 6-10~MeV SEP proton intensities
    observed during a solar eruption on Nov 3--4, 2011, at SOHO (top
    left panel), STEREO~B (top right) and STEREO~A (bottom left)
    spacecraft. The blue and red curves show the 10~MeV proton
    intensities simulated using the FP and FP+FLRW models,
    respectively, using an impulsive injection at flare maximum,
    $t_0=$22:41-500s UT on Nov 3 (dashed vertical line) at the flare
    longitude. The spacecraft locations are shown in the bottom right
  panel by filled green circle, yellow triangle, and blue square
for
  SOHO, STEREO~B, and STEREO~A, respectively. The blue arrow denotes
  the SEP injection longitude, the thick solid line and red circle are
  the same as in Fig.~\ref{fig:parker}, and the dashed curves show
  the Parker spiral field-lines connecting the spacecraft to the solar
  surface.}\label{fig:nov2011}
\end{figure*}

To compare the models with observed SEP events, we show in
Fig.~\ref{fig:nov2011} the time evolution of the simulated intensities
at three different longitudes for the two models, compared to an event
that took place during November 3--4, 2011. \citet{Gomez-Herrero2015}
have analysed this event using a 1D transport model, which, based on
anisotropy observations, suggests a source region with 270$^\circ$
longitudinal extent. However, they found no supporting evidence for
such a wide source and suggested a disturbed interplanetary
magnetic field as one of the possible causes for the event extent.

The SEP event is related to a solar eruption at longitude $50^\circ$
east of the STEREO~B (STB) spacecraft, with an M4.7 flare maximum
at $t_{\mathrm{max}}=$22:41~UT on November~3 at N20E62 from
Earth's point of view \citep{Richardson2014,Gomez-Herrero2015}. Using
the longitude angle $\phi$ as defined in Fig.~\ref{fig:parker},
where the particle source is at $\phi=0^\circ$, STB was at longitude
$\phi_{\mathrm{STB}}=50^\circ$, (see lower right panel of
Fig.~\ref{fig:nov2011}).  In this coordinate system, the SEP source
is connected along the Parker field line (solid black spiral curve in the
lower right panel of Fig.~\ref{fig:nov2011}) to longitude
$\phi=-62^\circ$ at radial distance of 1~AU (red circle).  During the
event, the STEREO~A (STA) and SOHO spacecraft were located at
longitudes $\phi_{\mathrm{STA}}=-103^\circ$ and
$\phi_{\mathrm{SOHO}}=152^\circ$, respectively.

Observed intensities of 6--10~MeV protons are shown in
Fig.~\ref{fig:nov2011} with dashed green curves. They were obtained
with ERNE onboard
SOHO \citep{Torsti1995} (top left panel), LET onboard STB
\citep{STEREOLET} (top right), and the LET instrument onboard STA (bottom
left). The blue and red curves in Fig.~\ref{fig:nov2011} show the
intensities from the FP and FP+FLRW models, respectively, for
the injection function $Q$ with
$t_0=t_{\mathrm{max}}-500 \mathrm{ s}$, (subtracting 500
  seconds of light-travelling time from Sun to 1~AU). The injection
intensity is fitted so that the highest intensities of the models
match the observed SEP highest intensity at STA.

The SEP event was best observed with LET/STA, located $41^\circ$
clockwise from the well-connected longitude of $\phi=-62^\circ$. The
6--10~MeV proton intensity increased by 5 orders of magnitude within a
few hours (bottom left panel of Fig.~\ref{fig:nov2011}). The event
was also observed by LET/STB and ERNE/SOHO at $112^\circ$ and
$146^\circ$, respectively, from the well-connected field-line. Even
with such poor connection, SEP intensities began to rise within an
hour of the observed SEP onset at STA, to rise by 3 orders of magnitude
above the pre-event background. It is therefore  clear that the SEPs had
rapid access to a wide range of longitudes during this solar event.

While we did not perform a full transport fitting in this report, our
study implies that the rapid access of SEPs during the November 3--4,
2011 event to the whole inner heliosphere can be achieved with our
FP+FLRW model even with a narrow source region. The FP+FLRW-modelled
particles are observed within an hour at the longitudes corresponding
to the November 3, 2011 spacecraft locations. The FP model can
reproduce the rapid onset of the SEP event at the STA spacecraft, with
a delay of half an hour compared to the FP+FLRW model. It is not
capable of reproducing the SOHO or STB observations, however, because
it fails to spread particles to these longitudes to intensity levels
above the pre-event background during the first ten hours of the
event.

\begin{figure}
  \includegraphics[width=\columnwidth]{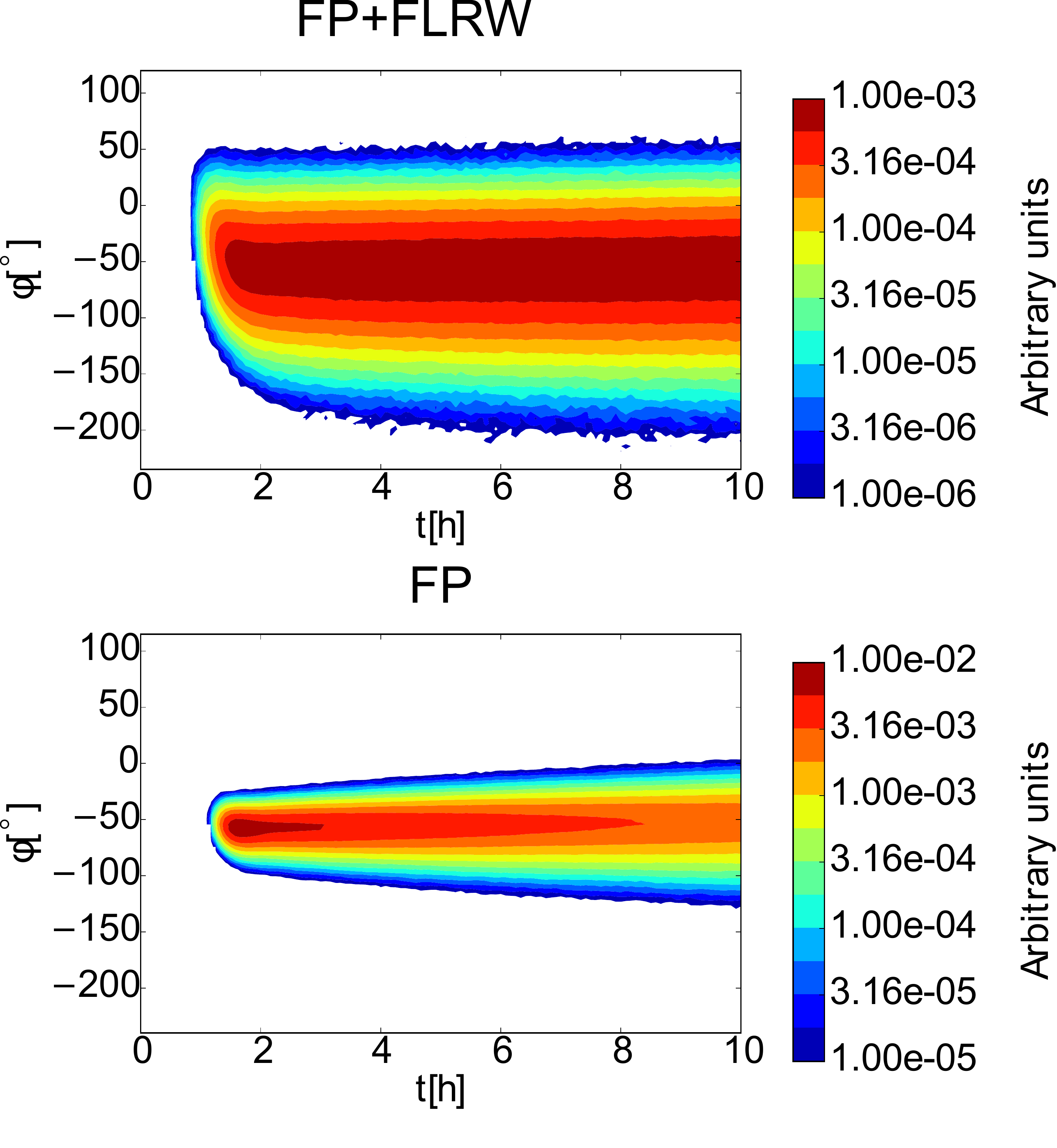}

\caption{Contour plot of particle density in arbitrary units as a
  function of time and longitude at 1 AU radial distance from the Sun
  (red circle in Figs. 1 and 2). The top panel shows the FP+FLRW model
  result, while the bottom panel shows results for the FP model.}\label{fig:phitime}
\end{figure}

The inability of the FP model to spread particles to the
  longitudes of the SOHO and STB spacecraft within the first ten hours
  is shown in Fig.~\ref{fig:phitime}, where we show the
evolution of the simulated SEP event as it would have been measured at 1~AU at
different heliographic longitudes, with $t_0=0$. For the FP model the intensity falls very sharply as a
  function of longitude, explaining the failure of the FP model in
  Fig.~\ref{fig:nov2011} to describe the November 2011 event without
  resorting to extremely wide source regions
  \citep{Gomez-Herrero2015}. As shown in Fig.~\ref{fig:phitime},
the SEPs spread quickly for the FP+FLRW model, and the SEP onset
within a wide range of longitudes, between $\phi=-150^\circ$ and
$50^\circ$, is delayed only by an hour compared to the onset at the
well-connected longitude, $\phi=-62^\circ$. Such a rapid spreading of
first particles is a key feature of observations: the earliest SEP
onsets at $100^\circ$ from the flare longitude are delayed only an
hour compared to onsets at best-connected longitude
\citep{Richardson2014}.

\begin{figure}
  \begin{center}
    \includegraphics[width=\columnwidth]{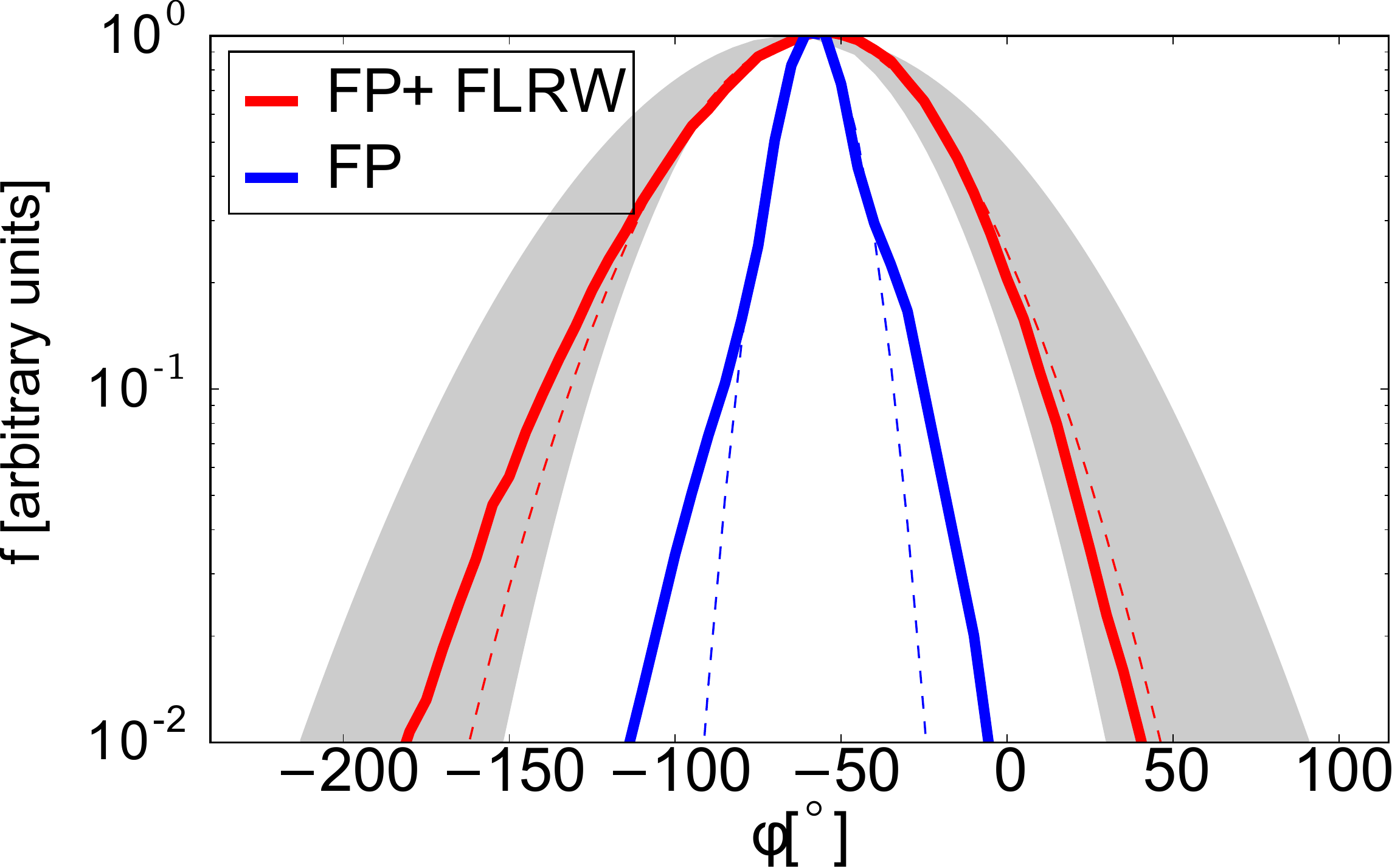}
  \end{center}
  \caption{Peak intensity at 1 AU as a function of
  longitude for the FP (blue solid curve) and FP+FLRW (red solid
  curve) models. The dashed lines show the fits of Gaussian profiles
  to the peak intensities, with $\sigma=11^\circ$ for the FP model,
  and $\sigma=34^\circ$ for the FP+FLRW model. The grey area depicts
  the range of Gaussian widths of $30-50^\circ$ obtained from
  observations.}\label{fig:peakdistr}
\end{figure}

The longitudinal extent of an SEP event is frequently described by
means of a Gaussian fit, $f(\phi)=f_0
\exp\{-(\phi-\phi_0)^2/(2\sigma^2)\}$, to the peak SEP intensities
observed by spacecraft at different longitudes. Using STEREO and
near-Earth spacecraft, several studies have reported values of
$\sigma=30^\circ-50^\circ$ to best represent the range of SEP event
widths
\citep{Lario2006,Lario2013,Wiedenbeck2013,Dresing2014,Richardson2014}. To
compare our models to observations, we performed a Gaussian fit to the
peak intensities of the simulated SEP events at 1~AU for the two
models presented in this work. The longitudinal distributions and the
corresponding fits are shown in Fig.~\ref{fig:peakdistr} with
solid and dashed curves, respectively. The grey area depicts the
observational range of $\sigma=30^\circ-50^\circ$. The FP+FLRW model
(red curve) can reproduce the observed SEP event widths, with width
$\sigma=34^\circ$, while the FP model (blue curve) can not; it
has a
very narrow SEP width, with only $\sigma=11^\circ$. This
  strong difference between the FP+FLRW and FP models is consistent
  with \citet{LaEa2013b}, who found for Cartesian geometry that at
  early times the mean square width in full-orbit simulations exceeded
  the FP model result by an order of magnitude and that an FP+FLRW model
  reprocudes the behaviour seen in full-orbit simulations well.

Our results show the importance of modelling the early phase using a
proper physical description. Both the FP and FP+FLRW aim to
  describe the effect of the turbulence on the particles, including
  the cross-field spreading that is due to field-line meandering. In the FP
  model, the field-line meandering is included only
  time-asymptotically using a diffusion description. However, as shown
  by \citet{LaEa2013b}, the initial spreading of particles is fast and
  non-diffusive, relaxing to diffusive behaviour only at later
  times. This is accounted for in our new FP+FLRW model and results in
  particles reaching much broader longitudinal extents than predicted
  by an FP model using the same turbulence parameters.

Propagation along meandering field lines implies that similar event
  extents should also be reached with other particle species.
This is indeed found in observations: \citet{Lario2013} and
\citet{Richardson2014} have found a similar range of peak intensity
widths for electrons and protons at different energies, and
\citet{Wiedenbeck2013} and \citet{Cohen2014} for heavy
ions during impulsive events.
While these observations do not
exclude other SEP-spreading mechanisms, such as transport in the
corona, symphatetic flaring and extended particle sources \citep[see,
e.g.,][]{Wiedenbeck2013,Gomez-Herrero2015}, the effect of
interplanetary field-line meandering offers a direct explanation for
the SEP event widths that are observed to be independent of energy and
particle species and the gradual or impulsive
  classification. The influence of the source height, size, and
  different interplanetary conditions on the resulting SEP event
  extent will be explored in a future full parametric study. We will
  also address the varying degrees of observed SEP anisotropies, which
  typically indicate cross-field transport in the interplanetary space
  \citep{Dresing2014}, but in some cases indicate rapid access of
  particles throughout the heliosphere \citep{Gomez-Herrero2015}.


\section{Conclusions}\label{sec:conclusions}

We developed a new SEP transport model that takes the non-diffusive propagation of SEPs early in the event
history into account for a Parker spiral geometry. We showed that the early onset
of SEPs over a wide range of longitudes can be explained by field-line
random walk and requires an SEP transport model that properly
describes the non-diffusive early phase of SEP cross-field
propagation. Our FP+FLRW model is the first model that is capable of
reproducing the observed fast access of SEPs to distant
  longitudes, when the particle and field-line diffusion coefficients
are consistently derived from an interplanetary turbulence
model. When the FLRW is not included (in the FP model), a much
  narrower cross-field extent of the SEP event is produced.  We
conclude that introducing field-line wandering into SEP modelling has
the potential of resolving the problem of fast access of SEPs to
  a wide range of longitudes.

\appendix

\section{Stochastic differential equations in the FP and FP+FLRW models}\label{sec:stoch-diff-equat}

The method of stochastic differential equations (SDEs) is a tool for solving
Fokker-Planck-type transport equations. Instead of solving this equation
directly, as is the case when applying finite-difference methods,
for example, the
SDEs trace so-called pseudo-particles, that is, phase-space elements, through
the phase-space. These pseudo-particles obey the corresponding Langevin equation. The
simplest example of a transport equation is
%
\begin{equation}
\pder{f}{t} = -v\pder{f}{x} + D\ppder{f}{x},
\end{equation}
with the Langevin equation
\begin{equation}
dx = v\>dt + \sqrt{2 D\,dt}\> W_t ,\label{eq:Langevin}
\end{equation}
 where $f(x,t)$ is the distribution function that depends on
  space $x$ and time $t$. $v$ and $D$ denote velocity and diffusion
  coefficient, $dt$ is the time step and and $W_t$ a Gaussian
  random number with zero mean and unit variance. The solution of the
  entire equation is obtained by binning the results for an
  appropriate number of pseudo-particles. Although these phase-space
  elements represent the behaviour of real particles in most
  cases quite well, they should not be confused with the latter. In contrast to
  the traditional approach, the SDE method also allows solving the
  transport equation, with a different Langevin equation,  backwards in time. For details on these points and on the method in
  general, we refer to \citet{Kopp2012} and the literature cited
  there.

In the present case we solved the transport equation, Eq.~(\ref{eq:FPE}) under the
assumption of a pure radial and constant solar wind
($\mathbf{V}_{sw}=V_{sw}\mathbf{e}_r$) and a vanishing latitudinal
component of the magnetic field, which in particular significantly reduces the Skilling term $R$ (the term in curved brackets in the second
line of Eq. (\ref{eq:FPE}) ). We first solved the equation in a
local Cartesian system with $z$ being the direction along the magnetic field,
and transformed the transport quantities back to the global system before the
actual integration was carried out \citep{Effenberger2012}. In this local system, the Langevin
equations read
\begin{eqnarray}
  dx &=& \left(\mathbf{u}_{\perp,x}+\pder{\kappa_{\perp,x}\,dt}{x}\right)\>dt+
  \sqrt{2 \kappa_{\perp,x}}\>W_{t,x},\nonumber\\
  dy &=& \left(\mathbf{u}_{\perp,y}+\pder{\kappa_{\perp,y}\,dt}{y}\right)\>dt+
  \sqrt{2 \kappa_{\perp,y}}\>W_{t,y},\nonumber\\
  d\mu &=& \frac{1}{2}\left(\frac{v}{L}+\mu R \right)(1-{\mu}^2)\>dt
  +\pder{D_{\mu\mu}}{\mu}\>dt+\sqrt{2 D_{\mu\mu}\,dt}\>W_{t,\mu}.\label{eq:LElocal}
\end{eqnarray}
 Here, $u_\parallel=V_{sw} b_r$, where $b_r$ is the radial
  component of the unit vector $\mathbf{\hat{b}}=\mathrm{e}_z$ along
  the local magnetic field line and $\mathbf{u}_\perp=V_{sw}
  (\mathbf{e}_r-b_r\mathbf{\hat{b}})$.  The two local unit vectors
  perpendicular to the magnetic field are $\mathbf{e}_x =
  \mathbf{e}_\theta \times \mathbf{e}_z$ and $\mathbf{e}_y =
  \mathbf{e}_z \times \mathbf{e}_x$, and the corresponding particle
  cross-field diffusion coefficients $\kappa_{\perp,x}$ and
  $\kappa_{\perp,x}$, respectively. For the FP model,
  $\mathbf{\hat{b}}$ is the unit vector along the local Parker field
  direction, whereas for the FP+FLRW model the pseudo-particle is
  first propagated (virtually) along the meandering field line, and
  $\mathbf{\hat{b}}$ itself is computed through the difference between the
  old and new points.

The transport coefficients in Eq. (\ref{eq:LElocal}) are
  transformed into the global system by means of the tensor
  $T_{\alpha\beta}=e_{\alpha,\beta}$, so that the (spatial) diffusion
  tensor becomes non-diagonal. In this case, the square root of a
  tensor has to be computed, see \citet{Kopp2012}.

It is important to note that calculating the meandering
  field line using Eq.~(\ref{eq:fldiff}) is performed before the
  particle is injected into the meandering field line. The
  particle cross-field propagation, while diffusive relative to the
  meandering field line, is therefore not diffusive relative to the mean
  magnetic field direction until the cross-field distance that
is due to
  integration of Eq.~(\ref{eq:LElocal}) is comparable to that due to
  the integration of Eq.~(\ref{eq:fldiff}). At early times, the
  perpendicular pseudo-particle propagation is therefore only weakly stochastic
  with respect to the mean magnetic field, making our model consistent
  with the results presented by \citet{LaEa2013b}.

\begin{acknowledgements} 
  We acknowledge the STEREO/IMPACT team for providing the data used in
  this study. The SOHO/ERNE data were obtained from the Space Research
  Laboratory, University of Turku (http://www.srl.utu.fi/erne\_data/).
  TL and SD acknowledge support from the UK Science and Technology
  Facilities Council (STFC) (grants ST/J001341/1 and ST/M00760X/1), MM
  from the European Commission FP7 Project COMESEP (263252), and FE
  from NASA grant NNX14AG03G.  The contribution of AK benefitted from
  financial support through project He 3279/15-1, funded by the
  Deutsche Forschungsgmeinschaft (DFG), at the CAU Kiel, where large
  parts of this work where carried out. TL, FE and SD acknowledge support
  from the International Space Science Institute as part of
  international team 297. We thank Horst Fichtner for helpful
  discussions. Access to the University of Central Lancashire's High
  Performance Computing Facility is gratefully acknowledged.
\end{acknowledgements}


\clearpage

\end{document}